\documentclass[runningheads, a4paper]{llncs}
 
\usepackage[T1]{fontenc}

\usepackage{graphicx}
\usepackage{algorithm}
\usepackage[noend]{algpseudocode}
\usepackage{algpseudocode}
\usepackage{amsmath, amsfonts}
\usepackage{nicematrix}
\usepackage{tikz}
\usepackage{tablefootnote}
\usepackage[outline]{contour}
\contourlength{1.2pt}

\usepackage{multirow}

\usepackage[available,functional]{usenixbadges}

\newcommand{\mathO}[1]{$\mathcal{O}({#1})$}
\newcommand{\eq}[1]{\textsc{eq}}

\usepackage{array}
\newcolumntype{P}[1]{>{\raggedright\arraybackslash}p{#1}}
\newcolumntype{C}[1]{>{\centering\arraybackslash}p{#1}}
\newcolumntype{H}{>{\setbox0=\hbox\bgroup}c<{\egroup}@{}}

\usepackage{booktabs}
\usepackage{subcaption}

\usepackage{listings}
 
\usepackage{varwidth}
\usepackage{hyperref}
\usepackage[utf8]{inputenc}
\usepackage[T1]{fontenc}
\usepackage{varioref}
\usepackage{siunitx}
 
\usepackage{tabularx}

\usepackage{arydshln}
\usepackage{nicematrix,tikz}
\usetikzlibrary{fit}

\usepackage[nottoc]{tocbibind}
\usepackage[numbers]{natbib}

\usepackage{xcolor}
\usepackage{wasysym}

\usepackage{caption}
\usepackage{subcaption}

\ifdefined\blindsubmission
    \def\LVS{{Levenshtein-trinary} }
    \def\LVSShort{{LS-3} }
\else
    \def\LVS{{Leuvenshtein}}
    \def\LVSShort{{\textsc{Lvs}} }
\fi

\usepackage{todonotes}

\newcommand*\Let[2]{\State #1 $\gets$ #2}

\newcommand{\tss}{\textsuperscript}
\newcommand{\deltah}{\ensuremath{\Delta h} }
\newcommand{\deltav}{\ensuremath{\Delta v} }
\newcommand{\deltahin}{\ensuremath{\Delta h_{in}} }
\newcommand{\deltahout}{\ensuremath{\Delta h_{out}} }\newcommand{\deltavin}{\ensuremath{\Delta v_{in}} }\newcommand{\deltavout}{\ensuremath{\Delta v_{out}} }

\begin{document}
 
\title{\LVS: Efficient FHE-based Edit Distance Computation with Single Bootstrap per Cell}

\author{Wouter Legiest\inst{1}, Jan-Pieter D'Anvers\inst{1}, Bojan Spasic\inst{2}, Nam-Luc Tran\inst{2} and Ingrid Verbauwhede\inst{1}}
\institute{COSIC, KU Leuven \\ \email{firstname.lastname@esat.kuleuven.be}\\ \and Society for Worldwide Interbank Financial Telecommunication (Swift)}

\authorrunning{Legiest et al.}

\maketitle

\begin{abstract}
    This paper presents a novel approach to calculating the Levenshtein (edit) distance within the framework of Fully Homomorphic Encryption (FHE), specifically targeting third-generation schemes like TFHE. Edit distance computations are essential in applications across finance and genomics, such as DNA sequence alignment. We introduce an optimised algorithm that significantly reduces the cost of edit distance calculations called \LVS{}. This algorithm specifically reduces the number of programmable bootstraps (PBS) needed per cell of the calculation, lowering it from approximately 94 operations -- required by the conventional Wagner-Fisher algorithm -- to just 1. Additionally, we propose an efficient method for performing equality checks on characters, reducing ASCII character comparisons to only 2 PBS operations. Finally, we explore the potential for further performance improvements by utilising preprocessing when one of the input strings is unencrypted. Our \LVS{} achieves up to $278\times$ faster performance compared to the best available TFHE implementation and up to $39\times$ faster than an optimised implementation of the Wagner-Fisher algorithm. Moreover, when offline preprocessing is possible due to the presence of one unencrypted input on the server side, an additional $3\times$ speedup can be achieved.
\end{abstract}

\section{Introduction}

The past 20 years have seen a major evolution of the global financial system. Financial crises, geopolitical events and economic growth have deeply impacted the direction that banking regulations have taken. One of the major policy shifts is in the direction of increasing transparency and sharing information between financial institutions. For instance, the G20 set targets for cross-border payments \cite{FSB2021} formulating objectives for enhancing cost, speed, financial inclusion and transparency in an effort to guarantee efficiency and seamlessness of an interconnected financial system. In line with the PSD2 directive \cite{PSD2}, which initiated the open-banking initiative in Europe, the EU has recently proposed the Financial Data Access framework which will grant consumers and SMEs to authorise third parties to access their data held by financial institutions. Information sharing among financial institutions is seen as paramount in the fight against financial crime and money laundering \cite{Maxwell2017}, and is also expected to drive GDP gains of major economies \cite{EC2022}.
In the previously mentioned initiatives, success can only be achieved if trust is built among all the actors. Trust can only be built if security and privacy are guaranteed in the exchange of information. While the directives are clear, the means to achieve a successful implementation are left to the actors proposing the services and products. 

Considering the recent legislative proposal to make Euro payments instant \cite{EC2022}, there is an obligation for payment providers to verify the match between the bank account number and the name of the beneficiary provided by the payer, as well as to alert the payer of possible mistakes or suspected fraud before the payment is made. In such applications, string similarity calculations are ubiquitous to provide robustness against spelling errors \cite{Alkhalili2021}. One example is the edit distance, which calculates the minimum number of edits between two given strings.

Recently, technologies enabling computation on encrypted data, namely fully homomorphic encryption (FHE) have become more practical. Informally, FHE is an encryption scheme that enables a data owner to securely outsource computation on their data to an untrusted processing party, whereby the processing party computes over encrypted data and stays oblivious of the data and the computed result. The utility of FHE comes at a performance price, which can sometimes be prohibitive for time-critical applications. However, the recent advances in software \cite{Chillotti2020} and hardware \cite{Geelen2023, CCS:vDTV23} implementation of the underlying FHE algorithms show promising performance results, encouraging the practitioners to start including FHE in production.

FHE schemes fall into two main categories: second-generation schemes like BGV~\cite{DBLP:journals/toct/BrakerskiGV14}, BFV~\cite{DBLP:journals/iacr/FanV12}, and CKKS~\cite{DBLP:conf/asiacrypt/CheonKKS17} support parallel computations on batched ciphertexts but have larger ciphertexts and slower bootstrapping. Third-generation schemes like TFHE~\cite{tfhe} prioritise speed with smaller ciphertexts and faster bootstrapping, though they work on small, individual messages and require bootstrapping for nearly every operation. TFHE also allows any function to be applied `for free' during bootstrapping, making it ideal for fast, logic-based encrypted computations.

In the context of string matching for financial applications, FHE could pose as an important enabler \cite{Maxwell2021}. Instead of physically sharing their customer information, parties could compute the desired outcome of the matching operation avoiding data sharing in clear altogether. The institution sending the payment would encrypt the transaction data using a suitable FHE scheme, and send it to a third party which would compute the desired matching score in the encrypted domain and return the encrypted result. This result (and any intermediate variable) can only be decrypted by the payer institution. In the process, according to the principles of FHE, the third party provably does not learn anything about the transaction data, and the institution sending the payment does not learn anything about the customers of the institution receiving the payment. 

Another interesting application of approximate string matching is in secure and privacy-preserving DNA analysis. Approximate string matching is essential in DNA analysis, where genetic sequences must be matched while allowing for slight discrepancies due to mutations. This flexibility is vital for detecting similar sequences that may vary because of natural mutations or sequencing errors. 

Cheon et al. \cite{CheonKL15} provided the first edit distance algorithm in the context of somewhat homomorphic encryption. They develop both equality check and min functions and use this to build up edit distance calculation. They also give a thorough analysis of the homomorphic depth of their solution. Their methods were later generalised by Vanegas et al.~\cite{VanegasCA23}, elaborated for an MPC context.
Later, Aziz et al. \cite{Aziz2017}, Asharov et al. \cite{PoPETS:AHLR18} and Zheng et al. \cite{8927885} proposed an approximation of the edit distance for genome analysis for fully homomorphic encryption. All the above techniques are based on second-generation FHE schemes and are built around arithmetic ciphertexts. 

Edit distance calculations for third-generation FHE schemes are less well-researched. Recently, ZAMA showed an edit distance calculation for TFHE as a demonstration of the concrete compiler~\cite{Concrete}. This demonstrator is based on high-level code (i.e., Python) that is transformed to TFHE by the concrete compiler, and the implementation uses a recursive definition of the edit distance.

\subsection{Our contribution}
In this paper, we develop a new edit distance algorithm for third-generation FHE schemes. We develop a new algorithm adapted to TFHE, which we call \LVS, and use this to show that the properties of the programmable bootstrapping play very well with edit distance calculations. The main ideas of our implementation are: 
\begin{enumerate}
    \item \textbf{Small representations}: We use differential values that represent the differences between intermediate results, instead of working on the intermediate results themselves. This reduces the size of the intermediate variables, leading to a smaller representation and more efficient calculations, significantly reducing the PBS costs. The size of our intermediate representations is small enough to fit in one ciphertext encoding 4 bits.
    \item \textbf{Re-using the programmable bootstrap}: In each iteration of the \LVS{} algorithm we have to produce two output values (i.e., the horizontal and vertical differential values). We show that one can rewrite the equations so that both output values can be computed with the same non-linear parts, which as a result means that they only differ by a (cheap) addition. As the main cost is in the non-linear part, which needs to be done using costly programmable bootstrapping, this technique reduces the calculation cost by roughly half.
    \item \textbf{Non-linear calculation in only one lookup}: During our calculation we have to compute the minimum of three inputs. A common strategy would be to do two bivariate lookups that each take two inputs. To enable the calculation of the non-linear part in only 1 programmable bootstrap, we propose a denser packing of the inputs. The input to our non-linear part has a total of $3\times3\times2 = 18$ values, while our 4-bit programmable bootstrap only allows a 16-value function. By adapting the non-linear function to start and end with zeros, we enable a larger effective lookup that can accommodate the full 18 values, saving another factor two in PBS.
\end{enumerate}

Our resulting \LVS{} algorithm\footnote{ A Rust implementation is available on \href{https://github.com/KULeuven-COSIC/leuvenshtein}{https://github.com/KULeuven-COSIC/leuvenshtein} and on \href{https://zenodo.org/records/15638825}{https://zenodo.org/records/15638825}} requires $94\times$ less PBS compared to a textbook Wagner-Fischer implementation, and $16\times$ compared to a bitsliced implementation (i.e., from Myers~\cite{Myers99}), excluding the equality calculations. 

Our second contribution is an optimised equality check implementation that uses significantly fewer programmable bootstraps (PBS). This method allows us to encode characters more optimally, reducing the number of ciphertexts required by half. More specifically, using our method, we are able to do an equality test on 7-bit ASCII strings in 2 PBS, instead of the standard 5 PBS as would be used by a standard equality check (as for example implemented in TFHE-rs).

A third contribution looks at preprocessing to reduce the (online) running cost. As our improved edit distance calculation only requires 1 PBS per edit distance, the main cost of the algorithm sits in the equality calculation. In case one of the input strings is unencrypted, we show that one can do a precalculation where each encrypted string is compared to each letter of the alphabet and all the results are stored in a lookup table. During the edit distance calculation one then only has to perform an (unencrypted) lookup to select the relevant equality value. This technique is useful when the encrypted string is known in advance, or when the alphabet is small compared to the string lengths. 

Combining these contributions, our implementation of the Levenshtein distance for ASCII inputs achieves a speedup of up to $278\times$ over the best available implementation, and a factor $40\times$ over our own state-of-the-art Wagner-Fisher implementation. In case of one unencrypted input, in some instances a further $3\times$ speedup is possible due to our improved preprocessing.

\section{Preliminaries}

In this section, we will first introduce the TFHE homomorphic encryption scheme. Then, we will define edit distances and specifically the Levenshtein distance, including the most relevant algorithms to calculate it.

\subsection{Notation}

For the rest of this work, we will compare two strings $a_{1..m}, b_{1..n}$, with lengths of $m$ and $n$ characters, respectively. All characters of the strings come from an alphabet $\Sigma$, and with $|\Sigma|$ we denote the number of characters in the alphabet. The $i$\textsuperscript{th} character of a string will be denoted as $a_i$. When the alphabet size is larger than the plaintext size, characters are represented through multiple symbols that are encrypted individually, denoted with $a_i^{(j)}$ for the $j$\textsuperscript{th} part of the $i$\textsuperscript{th} character of string $a$. For example, an 8-bit character can be split into two 4-bit symbols, $a^{(1)}$ and $a^{(2)}$. 

\subsection{TFHE}

Fully homomorphic encryption (FHE) enables computations to be carried out on encrypted data. This paper will focus on FHE schemes with programmable bootstrapping, specifically the Torus Fully Homomorphic Encryption (TFHE) scheme~\cite{tfhe}. We will provide a high-level introduction to TFHE; for more details, we refer to~\cite{tfhe, TCHES:Joye22}. 

In TFHE, a ciphertext typically holds 1-4 bits of plaintext while allowing for linear operations such as addition, subtraction, and multiplication with a small unencrypted value at a relatively low cost. However, these operations increase the noise in the ciphertexts. Once a certain number of operations have been performed, a noise reduction procedure called bootstrapping becomes necessary. Bootstrapping resets the noise, enabling further computations, but it is significantly more costly than linear operations. It is possible to chain ciphertexts together to encrypt larger integers. %, as large as \SI{256}{\bit}. JP: It can be larger than 256. So this sentence is irrelevant

One key advantage of TFHE is its capability to apply any lookup table (LUT) to the ciphertext without incurring any cost during bootstrapping. This process, known as programmable bootstrapping (PBS), facilitates executing highly non-linear functions on encrypted data. An example of such a LUT is shown in Table~\vref{table_lutrelu}.

\begin{figure}
\centering
    \includegraphics[width=0.6\textwidth]{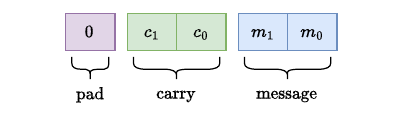}
    \caption{Subdivision of a 5-bit plaintext in 2-bit message and carry space for a TFHE ciphertext}
    \label{fig_tfhe_space}
\end{figure}

To illustrate, consider two 2-bit encrypted messages, $x$ and $y$, both encrypted in a 4-bit plaintext space (Figure~\vref{fig_tfhe_space}). To compare them, we first compute $x + (y \ll 2) \equiv (x + 4 \cdot y)$, resulting in a 4-bit value. The two least significant bits represent $x$, and the two most significant bits represent $y$. Then, we use a lookup table to check if the first 2 bits of the input equal the last 2 bits.

In more detail, the plaintext space is typically divided into message bits (the least significant bits of the plaintext space), carry bits (in the middle), and a padding bit (the most significant bit). The message bits represent plaintext values after encryption or bootstrapping. In contrast, the carry bits, initially zero, are filled after linear operations are performed (e.g. the $x + (y \ll 2)$ operation as described above). The padding bit is typically kept at zero to simplify the application of a LUT during programmable bootstrapping.

In more advanced scenarios, one can make an abstraction of the plaintext and carry space and use the entire plaintext space without the message-carry division. However, in this scenario, one should generally ensure that the padding bit remains zero to ensure proper LUT lookups during programmable bootstrapping.
\begin{table}[]
\centering
\caption{LUT Table for function $f(x) = x - 4$ in a 5-bit plaintext space. The right side can not be chosen, as it is the negative (mod 16) of the left side.}
\label{table_lutrelu}
\begin{tabular}{@{}rccrccrccrc@{}}
\toprule
\multicolumn{1}{c}{$x$}     & Output &  & \multicolumn{1}{c}{$x$}    & Output &  & \multicolumn{1}{c}{$x$}     & Output &  & \multicolumn{1}{c}{$x$}     & Output \\ \cmidrule(r){0-2} \cmidrule(lr){3-5} \cmidrule(lr){6-8} \cmidrule(l){9-11} 
$0 \; (0\,0000)$ & 12  &  & $8 \; (0\,1000)$ & 4   &  & $16 \; (\textbf{1}\,0000)$ & 4   &  & $24 \; (\textbf{1}\,1000)$ & 12  \\
$1 \; (0\,0001)$ & 13  &  & $9 \; (0\,1001)$ & 5   &  & $17 \; (\textbf{1}\,0001)$ & 3   &  & $25 \; (\textbf{1}\,1001)$ & 11  \\
$2 \; (0\,0010)$ & 14  &  & $10 \; (0\,1010)$ & 6   &  & $18 \; (\textbf{1}\,0010)$ & 2   &  & $26 \; (\textbf{1}\,1010)$ & 10  \\
$3 \; (0\,0011)$ & 15  &  & $11 \; (0\,1011)$ & 7   &  & $19 \; (\textbf{1}\,0011)$ & 1   &  & $27 \; (\textbf{1}\,1011)$ & 9   \\
$4 \; (0\,0100)$ & 0   &  & $12 \; (0\,1100)$ & 8   &  & $20 \; (\textbf{1}\,0100)$ & 0   &  & $28 \; (\textbf{1}\,1100)$ & 8   \\
$5 \;(0\,0101)$ & 1   &  & $13 \; (0\,1101)$ & 9   &  & $21 \; (\textbf{1}\,0101)$ & 15  &  & $29 \; (\textbf{1}\,1101)$ & 7   \\
$6 \;(0\,0110)$ & 2   &  & $14 \; (0\,1110)$ & 10  &  & $22 \; (\textbf{1}\,0110)$ & 14  &  & $30 \; (\textbf{1}\,1110)$ & 6   \\
$7 \;(0\,0111)$ & 3   &  & $15 \; (0\,1111)$ & 11  &  & $23 \; (\textbf{1}\,0111)$ & 13  &  & $31 \; (\textbf{1}\,1111)$ & 5   \\ \bottomrule
\end{tabular}
\end{table}

More specifically, we can create the corresponding Lookup Table (LUT) for any arbitrary function as long as the padding bit is 0. Due to the nature of TFHE calculations, when the padding bit is 1, the lookup result will be the negative of the corresponding input with the padding bit 0. For example, if using a 5-bit plaintext space to evaluate function $f$ through a LUT, and we want to use the entire 5-bit, we must consider that for input values $x = 2^{5} / 2 = 16<x< 2^{5} = 32$, the function will become $f(x) = -f(x-16)$. This property is due to the negacyclic nature of the polynomials and is inherent in any current FHE scheme with PBS. 

Another critical aspect of FHE is that data-dependent branching (e.g., if, while statements) cannot be used due to its confidential nature. To evaluate a branch in FHE, all possible outcomes must be calculated. This means that ideally, programs need to be rewritten to avoid if statements, and if statements cannot be used to skip irrelevant parts of the execution. We will revisit this topic in our discussion of edit distance calculation algorithms.

In the following sections, we will use a parameter set with a plaintext size of 5 bits. In this set, the lowest 4 bits can be freely assigned, while the most significant bit is utilised for padding. This parameter set is commonly used in practice.

\subsection{Edit distance}
The \emph{edit distance} is a metric used to measure the similarity between two strings by calculating the number of edit operations needed to transform one string into another. It differs from the Hamming distance, which only considers the similarity of corresponding characters.
For example, the Hamming distance between \texttt{`abcdex'} and \texttt{`xabcde'} is six, while the edit distance is two (one insertion and one deletion).

The edit distance exists in various variants where each variant allows a different set of operations: the first two operations, \emph{`insertion'} and \emph{`deletion'}, correspond to the addition or removal of a character. The third operation is \emph{`substitution'}, which involves replacing one character with another. The fourth operation is \emph{`transposition'}, where two adjacent characters swap places. In this work we will focus on the Levenshtein distance, which considers the first three operations.

There are various versions of the edit distance. For example, costs can be assigned to each operation, allowing different weights to be applied. When all operations have a uniform unit cost, it is referred to as \emph{simple edit distance}. If non-unit costs are used, it is called \emph{general edit distance}.
In some cases, the goal is to find the exact value below a specific limit, and once that limit is exceeded, the exact value becomes unimportant. An \emph{approximated edit distance} can be used in such scenarios.

The popular Levenshtein distance is often interchangeably used with edit distance. 
 
These metrics are popular tools in (financial) fraud detection, DNA sequence comparison, calculating distances between matrix sequences~\cite{PayneHRL19,4603758,BEERNAERTS2019373}, and spell checkers.

\subsubsection{Calculating the edit distance}

The Levenshtein distance was originally obtained using a recursive definition. This definition was later converted to an executable algorithm, the Wagner-Fischer algorithm, using dynamic programming. Since then, more efficient variations have been proposed, improving time and space complexity. For a comprehensive overview of this plaintext algorithm, refer to the work of Navarro~\cite{Navarro01}.

Advanced algorithms, such as those based on the Four Russians Method~\cite{MP80}, Suffix trees~\cite{Knuth73}, or filtering~\cite{Uk92}, are not suitable for implementation in FHE due to their data-dependent assumptions or alphabet-specific data representations. Additionally, algorithms based on nondeterministic finite automaton (NFA)~\cite{Uk85b} will have the same complexity as the Wagner-Fischer algorithm in the encrypted domain. The representation of the automaton will have the same form as the $d$-matrix. Therefore, specific FHE-friendly optimisations are needed to speed up the calculations of the edit distance further.

\subsubsection{Wagner-Fischer}

The Wagner-Fischer algorithm~\cite{Vintsyuk1968SpeechDB, WagnerF74} uses dynamic programming to create a distance matrix (or `$d$-matrix').  %, Figure~\ref{fig_dmatrix}. 
Each element in the matrix represents the edit distance of the corresponding substrings up to that point in the matrix. For instance, $D[i,j] = ed(a_{1..i}, b_{1..j})$ corresponds to the edit distance of the first $i$ characters of string $a$ and the first $j$ characters of string $b$. Specifically, in the simple edit distance case, each value of the $d$-matrix is determined by the following equation:
\begin{equation}
\label{eq_wf}
    D[i,j] = 
    \begin{cases}    
    D[i-1,j-1] & \text{if}\ a_i = b_j \\
    1 + \min(D[i-1,j], D[i,j-1], D[i-1,j-1]) & \text{otherwise}.
    \end{cases}
\end{equation}

To calculate the next value, the algorithm uses three previously computed values. These dependencies make the algorithm difficult to parallelise. The original definition has a time and memory complexity of \mathO{n^2}. A simple optimisation is to reduce the space complexity to \mathO{n} by only storing some columns of the $d$-matrix. Examples of the $d$-matrix are given in Figure~\vref{fig_dmatrix}.

Since its definition, many variations have been proposed to optimise the calculation. They mostly rely on skipping parts of the calculations based on the alphabet or data-dependent intermediate values~\cite{Myers86, Ukkonen85a}. It is impractical to port these optimisations to the FHE domain, as we do not know the value of intermediate variables and can thus not do any data-dependent optimisations. 

\begin{figure}[!ht]
  \centering
      \begin{minipage}[b]{.32\textwidth}
  \centering
\begin{tabular}{llllllll}
                       &            & f          & r          & i          & d          & a          & y          \\ \cline{2-8} 
\multicolumn{1}{l|}{}  & \textbf{0} & 1          & 2          & 3          & 4          & 5          & 6          \\
\multicolumn{1}{l|}{m} & 1          & \textbf{1} & 2          & 3          & 4          & 5          & 6          \\
\multicolumn{1}{l|}{o} & 2          & 2          & \textbf{2} & 3          & 4          & 5          & 6          \\
\multicolumn{1}{l|}{n} & 3          & 3          & 3          & \textbf{3} & 4          & 5          & 6          \\
\multicolumn{1}{l|}{d} & 4          & 4          & 4          & 4          & \textbf{3} & 4          & 5          \\
\multicolumn{1}{l|}{a} & 5          & 5          & 5          & 5          & 4          & \textbf{3} & 4          \\
\multicolumn{1}{l|}{y} & 6          & 6          & 6          & 6          & 5          & 4          & \textbf{3}
\end{tabular}
    \end{minipage}
        \begin{minipage}[b]{.32\textwidth}
          \centering
    \begin{tabular}{llllll}
                       &   & x & a & b & c \\ \cline{2-6} 
\multicolumn{1}{l|}{}  & \textbf{0} & 1 & 2 & 3 & 4 \\
\multicolumn{1}{l|}{a} & 1 & \textbf{1} & \textbf{1} & 2 & 3 \\
\multicolumn{1}{l|}{b} & 2 & 2 & 2 & \textbf{1} & 2 \\
\multicolumn{1}{l|}{c} & 3 & 3 & 3 & 2 & \textbf{1} \\
\multicolumn{1}{l|}{x} & 4 & 3 & 4 & 3 & \textbf{2}
    \end{tabular}
 \end{minipage}

  \caption{$d$-matrix of the (Simple) Edit distances of $d(\texttt{`monday'},\texttt{`friday'}) = 3$ and $d(\texttt{`abcx'},\texttt{`xabc'}) = 2$}
  \label{fig_dmatrix}
\end{figure}

\subsubsection{Myers}
\label{sec:myers}

An alternative approach to Wagner-Fischer was proposed by Myers~\cite{Myers99}. This approach targets modern CPUs by rewriting the algorithm in terms of bits and optimising it for this lead. The main idea is to store the differential values (or differences between adjacent horizontal and vertical cells) in the $d$-matrix instead of absolute distances.

{\renewcommand{\arraystretch}{2.8}
\setlength{\tabcolsep}{14pt}
\begin{figure}
    \centering
\begin{minipage}[b]{.48\textwidth}
    \begin{NiceTabular}{|cccc}[first-row,first-col]

      & & S & I & T \\
    \cline{1-5}
      & 0 & 1 & 2 & 3  \\
    K $\quad\;\;$ & 1 & 1 & 2 & 3  \\
    I $\quad\;\;$ & 2 & 2 & 1 & 2  \\
    D $\quad\;\;$ & 3 & 3 & 2 & 2  \\
    % \hline
    \end{NiceTabular}
        % \caption{Edit distance calculation of $d(\texttt{'KID'},\texttt{'SIT'}) = 2$ through the Myers algorithm.}
        % \label{fig:enter-label}
\end{minipage} %
\begin{minipage}[b]{.48\textwidth}

    \centering
\begin{NiceTabular}{|cccc}[first-row,first-col]

  & & S & I & T \\
\cline{1-5}
  & \textcolor{white}{1} & \textcolor{white}{1} & \textcolor{white}{1} & \textcolor{white}{1}  \\
& \textcolor{white}{1} & \textcolor{white}{1} & \textcolor{white}{1} & \textcolor{white}{1}  \\
& \textcolor{white}{1} & \textcolor{white}{1} & \textcolor{white}{1} & \textcolor{white}{1}  \\
& \textcolor{white}{1} & \textcolor{white}{1} & \textcolor{white}{1} & \textcolor{white}{1}  \\
 
\CodeAfter
\begin{tikzpicture}
\begin{scope}[->,shorten < = 4pt, shorten > = 4pt]
\draw (1-1) --node [midway, above] {\contour{white}{\scriptsize +1}} (1-2) ;
\draw (1-2) --node [midway, above] {\contour{white}{\scriptsize+1}} (1-3) ;
\draw (1-3) --node [midway, above] {\contour{white}{\scriptsize+1}} (1-4) ;

\draw (2-1) --node [midway, above] {\contour{white}{\scriptsize0}} (2-2) ;
\draw (2-2) --node [midway, above] {\contour{white}{\scriptsize+1}} (2-3) ;
\draw (2-3) --node [midway, above] {\contour{white}{\scriptsize+1}} (2-4) ;

\draw (3-1) --node [midway, above] {\contour{white}{\scriptsize0}} (3-2) ;
\draw (3-2) --node [midway, above] {\contour{white}{\scriptsize-1}} (3-3) ;
\draw (3-3) --node [midway, above] {\contour{white}{\scriptsize+1}} (3-4) ;

\draw (4-1) --node [midway, above] {\contour{white}{\textbf{\scriptsize0}}} (4-2) ;
\draw (4-2) --node [midway, above] {\contour{white}{\textbf{\scriptsize-1}}} (4-3) ;
\draw (4-3) --node [midway, above] {\contour{white}{\textbf{\scriptsize 0}}} (4-4) ;

\draw (1-1) --node [midway, left] {\contour{white}{\textbf{\scriptsize+1}}} (2-1) ;
\draw (2-1) --node [midway, left] {\contour{white}{\textbf{\scriptsize+1}}} (3-1) ;
\draw (3-1) --node [midway, left] {\contour{white}{\textbf{\scriptsize+1}}} (4-1) ;

\draw (1-2) --node [midway, left] {\contour{white}{\scriptsize0}} (2-2) ;
\draw (2-2) --node [midway, left] {\contour{white}{\scriptsize+1}} (3-2) ;
\draw (3-2) --node [midway, left] {\contour{white}{\scriptsize+1}} (4-2) ;

\draw (1-3) --node [midway, left] {\contour{white}{\scriptsize0}} (2-3) ;
\draw (2-3) --node [midway, left] {\contour{white}{\scriptsize-1}} (3-3) ;
\draw (3-3) --node [midway, left] {\contour{white}{\scriptsize+1}} (4-3) ;

\draw (1-4) --node [midway, left] {\contour{white}{\scriptsize0}} (2-4) ;
\draw (2-4) --node [midway, left] {\contour{white}{\scriptsize-1}} (3-4) ;
\draw (3-4) --node [midway, left] {\contour{white}{\scriptsize0}} (4-4) ;

\end{scope}
\end{tikzpicture}
\end{NiceTabular}
    % \caption{Edit distance calculation of $d(\texttt{'KID'},\texttt{'SIT'}) = 2$ through the Myers algorithm.}
    % \label{fig:enter-label}
\end{minipage}
\caption{Edit distance calculation of $d(\texttt{`KID'},\texttt{`SIT'}) = 2$ through the Wagner-Fisher algorithm (left), where the absolute distances are calculated; and the Myers algorithm (right), which calculates the relative distances.}
        \label{fig_myers}
\end{figure}
}

In simple edit distance, each neighbouring value in the $d$-matrix can differ by at most one (see Eq.~\ref{eq_wf}). The Myers algorithm takes advantage of this by only representing the horizontal and vertical differences between neighbouring cells in the $d$-matrix (Figure~\vref{fig_myers}). This allows Boolean logic to compute the distance, making it possible to better utilise hardware parallelisation, particularly on CPUs. Note that from these horizontal and vertical differences, it is straightforward to reconstruct any value in the $d$-matrix by choosing a path from the start to the targeted cell and summing the horizontal and vertical differences along this path.

The core of the algorithm is to store only the neighbour differences using ternary values $\{-1,0,1\}$, both in the vertical and horizontal direction:
\begin{align}
    \label{eq:delta}
    \begin{split}
    \Delta v[i,j] &= D[i,j] - D[i-1,j],  \\
    \Delta h[i,j] &= D[i,j] - D[i,j-1].
    \end{split}
\end{align}

\begin{figure}[!ht]
\centering
    \includegraphics[width=0.5\textwidth]{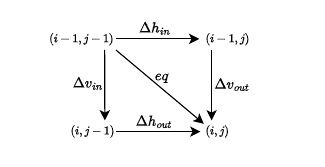}
    \caption{Myers cell~\cite{Myers99}}
    \label{fig_Myerscell}
\end{figure}

We can use the following equations to directly calculate the $\deltav$ and $\deltah$ values. When we examine a cell $(i,j)$, we can define $\deltavout$ and $\deltahout$ as the values $\Delta v[i,j]$ and $\Delta h[i,j]$. These are the values we aim to compute in this cell. The equations above can then be rephrased in terms of the equality $\textsc{eq}$ between the two relevant characters $a_i$ and $b_j$, and the previous values of $\deltav$ and $\deltah$ which we denote as $\deltavin$ and $\deltahin$. An overview of the input and output variables is given in \autoref{fig_Myerscell}.

We can transform \autoref{eq:delta} to calculate the outputs $\deltavout$ and $\deltahout$ of a single $d$-matrix cell, in function of the inputs $\textsc{eq}$, $\deltavin$, $\deltahin$:

\begin{align*}
    \Delta v_{out} &= \min\left(
        \begin{array}{l}
             1, \\
             \deltavin + 1 - \deltahin,  \\
             1 - \textsc{eq} - \deltahin
        \end{array} \right), \\
    \Delta h_{out} &= \min\left(
        \begin{array}{l}
             1, \\
             1 + \deltahin - \deltavin,  \\
             1 - \textsc{eq} - \deltavin
        \end{array} \right). 
\end{align*}

Building on this concept, it is possible to transform the entire Wagner-Fischer algorithm to use only Boolean operations and additions. For instance, they define two Boolean values to represent the ternary nature of the delta elements. 
By leveraging a $w$ bit CPU architecture, the algorithm achieves a time complexity of \mathO{\lceil m/w\rceil n}. However, the Myers algorithm does not necessarily translate well to a TFHE environment, where operations can be performed on multi-bit values. A detailed overview of the Myers algorithm is provided in~\cite{Myers99}.

\section{Encrypted Levenshtein}

Edit distance calculation generally involves two phases: equality checking and the main algorithm. The equality checking phase determines character equality between the input strings. The main algorithm then uses these equalities to compute the $d$-matrix (or differentials in the Myers approach) and determine the edit distance. This section will focus on the main algorithm, while \autoref{sec:equality} will focus on the equality checking phase. 

\subsection{Main algorithm}

This section will develop a new edit distance algorithm suited for the encrypted domain. Our algorithm improves the Myers approach detailed in \autoref{sec:myers} in the FHE case. As a reminder, the main idea of this approach is to calculate the differential values between two nodes in the edit distance calculation, as given in \autoref{fig_Myerscell}.

The Myers approach is designed for CPU-optimised operations, such as Boo\-le\-an operations and additions, by using bitslicing. This makes operations very efficient on CPU, but they do not necessarily translate to efficient operations in the TFHE domain. Our approach focuses on optimising for FHE in two steps:

First, our approach uses small multivalue operands (typically in the range $\{-1, 0, 1\}$) instead of binary values in Myers. For example, we represent the trinary $\deltahin$ and $\deltavin$ operators in one operand instead of splitting into binary positive and negative parts, as done by Myers. This limits the number of inputs and outputs that need to be handled and is efficient due to the native multivalue operations in TFHE.

Secondly, we rewrite the cell equations to allow the calculations in only one bootstrap. This means that $\deltavin$, $\deltahin$, and \textsc{eq} are given as inputs to the PBS, and $\deltavout$, $\deltahout$ are extracted as the outputs. To achieve this, we have to optimise the PBS to perform a lookup that is relevant for both the outputs $\deltavout$ and $\deltahout$. 
 
We then show that we can still perform this lookup in 1 PBS by carefully manipulating the PBS function.

\subsubsection{Combining the PBS calculations}

Using the Myers approach, our algorithm calculates two output values $\deltavout$ and $\deltahout$ for each cell, as explained in \autoref{fig_Myerscell}. 
 
The formulas to compute $\Delta v_{out}$, $\Delta h_{out}$ are shown below: 

\begin{align*}
    \Delta v_{out} &= \min\left(
        \begin{array}{l}
             1, \\
             \deltavin + 1 - \deltahin,  \\
             1 - \textsc{eq} - \deltahin
        \end{array} \right), \\
    \Delta h_{out} &= \min\left(
        \begin{array}{l}
             1, \\
             1 + \deltahin - \deltavin,  \\
             1 - \textsc{eq} - \deltavin
        \end{array} \right). 
\end{align*}

The first optimisation is to rewrite the equations to have a similar non-linear operation. We can rewrite both equations to: %formula's~\ref{mye_eq_1} and \ref{mye_eq_2} in Myers. 

\begin{align}
\begin{split}
    \Delta v_{out} &= \min(-\textsc{eq}, \Delta v_{in}, \Delta h_{in}) + (1 - \Delta h_{in}),  \\
    \Delta h_{out} &= \min(-\textsc{eq}, \Delta v_{in}, \Delta h_{in}) + (1 - \Delta v_{in}) . \label{mye_eq_2}
    \end{split}
\end{align}
In this form, we can focus on $\min(-\textsc{eq}, \Delta v_{in}, \Delta h_{in})$ in the PBS, and perform the $(1 - \Delta h_{in})$ operations using linear computations without bootstrapping at low cost. This optimisation reduces the calculation cost with approximately a factor 2.

\subsubsection{Extended lookups}
A standard approach to calculate the min function in \autoref{mye_eq_2} would be to do two bivariate lookups, which would first combine two inputs into the key $(\text{key} = \deltavin + 4 \cdot \deltahin)$, after which a PBS with relevant lookup table is performed on this key to calculate the function $\min(\deltavin, \deltahin)$. In the second phase, a similar min function is performed between the result of the first min function and $\textsc{eq}$. This approach requires two PBS lookups for each cell of the \LVS{} calculation. 

In this section we will reduce this further to one PBS per cell in the standard Levensthein case, by combining the calculation of both min functions. In this case, we have $\Delta v_{in}$, $\Delta h_{in} \in [-1, 0, 1]$ and $\textsc{eq} \in [0, 1]$. By combining the inputs to the min functions in a more dense way (e.g. $\deltavin + 3 \cdot \deltahin + 9 \cdot \textsc{eq}$) one could reduce the input size. However, even in the best case, this entails a $3\times3\times2=18$-entry lookup table for the min operation, while we only have a 16-entry lookup table available.

It is important to note the negacyclic nature of the TFHE-PBS lookup. In TFHE with a 4-bit message (and carry), one can construct any lookup table of 16 values (values 0 to 15). Lookups at values 16 to 31 will result in the negated value of the corresponding value at position $i-16$. This means that if we can place a zero lookup at position 0, and the value at position 16 is also 0, we can essentially extend the lookup table with one extra value. We can extend this as long as the $i$\tss{th} and $(i+16)$\tss{th} values are both 0.

When we rewrite the formulas for \deltavout and \deltahout as:
\begin{align*}
    \Delta v_{out} = \left\{1 + \min(-\textsc{eq}, \Delta v_{in}, \Delta h_{in}) \right\} - \Delta h_{in}, \\
    \Delta h_{out} = \left\{1 + \min(-\textsc{eq}, \Delta v_{in}, \Delta h_{in}) \right\} - \Delta v_{in} .
\end{align*}
The function between brackets returns mostly zeros during the programmable bootstrap. We will denote this value with $M$, or $M_{ij}$, denoting the $M$ value in the cell at location $i,j$. Combining this with the following key for the PBS lookup:
\begin{align}
    (\deltavin + 1) + 3 \cdot (1 + \deltahin) + 9 \cdot \textsc{eq} \label{eq_keyeq}
\end{align}
results in a lookup table of 18 values, which starts and ends with two zeros. This allows us to fit this lookup table into a 16-value TFHE lookup table, as detailed in \autoref{table:LUTmin}.

\begin{table}[]
\centering
\caption{LUT Table for function LUT$_{\text{min}}: 1 + \min(-\textsc{eq}, \Delta v_{in}, \Delta h_{in})$, with $x=(\deltavin + 1) + 3 \cdot (1 + \deltahin) + 9 \cdot \textsc{eq}$ in a 5-bit plaintext space. Note that the right side is the negative of the left side (mod 16) due to the negacyclic nature of the LUT.}
\label{table:LUTmin}
\begin{tabular}{@{}rccrccrccrc@{}}
\toprule
\multicolumn{1}{c}{$x$}    & Output &  & \multicolumn{1}{c}{$x$}    & Output &  & \multicolumn{1}{c}{$x$}    & Output &  & \multicolumn{1}{c}{$x$}    & Output \\ \cmidrule(r){0-2} \cmidrule(lr){3-5} \cmidrule(lr){6-8} \cmidrule(l){9-11} 
$0 \; (0\,0000)$ & 0  &  & $8 \; (0\,1000)$ & 1   &  & $16 \; (\textbf{1}\,0000)$ & 0   &  & $24 \; (\textbf{1}\,1000)$ & 15  \\
$1 \; (0\,0001)$ & 0  &  & $9 \; (0\,1001)$ & 0   &  & $17 \; (\textbf{1}\,0001)$ & 0   &  & $25 \; (\textbf{1}\,1001)$ & 0  \\
$2 \; (0\,0010)$ & 0  &  & $10 \; (0\,1010)$ & 0   &  & $18 \; (\textbf{1}\,0010)$ & 0   &  & $26 \; (\textbf{1}\,1010)$ & 0  \\
$3 \; (0\,0011)$ & 0  &  & $11 \; (0\,1011)$ & 0   &  & $19 \; (\textbf{1}\,0011)$ & 0   &  & $27 \; (\textbf{1}\,1011)$ & 0   \\
$4 \; (0\,0100)$ & 1   &  & $12 \; (0\,1100)$ & 0   &  & $20 \; (\textbf{1}\,0100)$ & 15   &  & $28 \; (\textbf{1}\,1100)$ & 0   \\
$5 \;(0\,0101)$ & 1  &  & $13 \; (0\,1101)$ & 0   &  & $21 \; (\textbf{1}\,0101)$ & 15  &  & $29 \; (\textbf{1}\,1101)$ & 0   \\
$6 \;(0\,0110)$ & 0   &  & $14 \; (0\,1110)$ & 0  &  & $22 \; (\textbf{1}\,0110)$ & 0  &  & $30 \; (\textbf{1}\,1110)$ & 0   \\
$7 \;(0\,0111)$ & 1   &  & $15 \; (0\,1111)$ & 0  &  & $23 \; (\textbf{1}\,0111)$ & 15  &  & $31 \; (\textbf{1}\,1111)$ & 0   \\ \bottomrule
 
\end{tabular}
\end{table}

\subsubsection{Limiting the noise growth}

Our approach yields the same result as the corresponding plaintext algorithms. However, since the computation is performed under FHE, there remains a small probability of error. This probability is typically kept extremely low for cryptographic security purposes (for instance, below $2^{-64}$) by carefully choosing parameters that control noise growth. In this subsection, we demonstrate that, with a few minor adaptations, our algorithm remains within the predefined noise bounds, thereby preserving the intended low failure probability. This failure probability is sufficiently low that it should not be noticeable by end-users.

In the previous discussion, we simplified the operations in two cases: costly non-linear bootstraps and cheap linear operations that do not require a bootstrap. In reality, there is a maximum number of linear operations that can be performed before a bootstrap is needed. In our case the calculation of the key can exceed this threshold when not taken into account. 

We will denote the variance of the noise of one ciphertext with $\epsilon^2_{PBS}$ (i.e., the noise of a ciphertext at bootstrap time when it has not been combined with another ciphertext). When adding $k$ independent ciphertexts, the noise will be equivalent to $k \cdot \epsilon^2_{PBS}$, while the multiplication of a ciphertext with $k$ results in a noise equivalent of $k^2 \cdot \epsilon^2_{PBS}$. These come from standard equations for the addition and multiplication of stochastic variables.

For the key calculation in \autoref{eq_keyeq}, one thus has a noise equivalent of:
\begin{align*}
    \epsilon^2_{key} = \epsilon^2_{\deltavin} + 9 \cdot \epsilon^2_{\deltahin} + 81 \cdot \epsilon^2_{\textsc{eq}_9} .
\end{align*}
Note that for now, we assume independence between the variables, which we will come back to later in this section.

A first trick to reduce the error is to include the factor 9 at the \textsc{eq} term: instead of calculating $\text{enc}(\textsc{eq})$ during the equality checking phase, we adapt the bootstrap to calculate $\text{enc}(9 \cdot \textsc{eq})$. Moreover, this gives us a small speedup since a scalar multiplication is avoided. We will denote the value $9 \cdot \textsc{eq}$ with $\textsc{eq}_9$. This reduces the noise of the key to:
\begin{align*}
    \epsilon^2_{key} = \epsilon^2_{\deltavin} + 9 \cdot \epsilon^2_{\deltahin} + \epsilon^2_{\textsc{eq}_9} 
    & = \epsilon^2_{\deltavin} + 9 \cdot \epsilon^2_{\deltahin} + \epsilon^2_{PBS} .
\end{align*}

A second thing to notice is that \deltavin and \deltahin themselves are calculated recursively. Denoting with $\Delta v_{i,j}$ the $\deltavout$ of cell $(i,j)$ (and similarly for $\deltahout$) and $M_{i,j} = D[i,j] - D[i-1,j-1]$, we get the following formulas:

\begin{equation*}
\begin{split}
    \deltavin =& \Delta v_{i,j-1} \\
    =& M_{i,j-1} - \Delta h_{i-1,j-1} \\
    =& M_{i,j-1} - M_{i-1,j-1} + \Delta v_{i-1,j-2} \\
    =& M_{i,j-1} - \pmb{M_{i-1,j-1}} + M_{i-1,j-2} - \pmb{M_{i-2,j-2}} + \cdots \\
    \deltahin =& M_{i-1,j} - \pmb{M_{i-1,j-1}} + M_{i-2,j-1} - \pmb{M_{i-2,j-2}} + \cdots
\end{split}
\end{equation*}

All $M$ are calculated using different inputs, and thus, their respective noise can be considered independent. However, the formulas of \deltavin and \deltahin have the same negative $M$ terms (in bold in the equation above), which increases the noise in the key:

\begin{equation*}
\begin{split}
    key &= \deltavin + 3 \cdot \deltahin + \textsc{eq}_9 +Cte \\
    &= (M_{i,j-1} - 4 \cdot M_{i-1,j-1} + 3 \cdot M_{i-1,j}) + \\
    & (M_{i-1,j-2} - 4 \cdot M_{i-2,j-2} + 3 \cdot M_{i-2,j-1}) + \\
    & \cdots + \\
    & \textsc{eq}_9 + Cte
\end{split}
\end{equation*}

or:
\begin{align*}
   \epsilon^2_{key} = N_H \cdot \epsilon^2_{PBS}  + 16 \cdot N_M \cdot \epsilon^2_{PBS} + 9 \cdot N_L \cdot \epsilon^2_{PBS} + \epsilon^2_{PBS}
\end{align*}
with $N_H$, $N_M$ and $N_L$ the number of respective $M$ terms (i.e., $N_H$ is the number of $M_{i,j-1}$-like terms, $N_M$ the number of $4 \cdot M_{i-1,j-1}$-like terms and $N_L$ the number of $3 \cdot M_{i-1,j}$-like terms). 

The noise can be easily reduced by changing the equation of the key as:
\begin{align}
    (1 - \deltavin) + 3 \cdot (1 + \deltahin) + 9 \cdot \textsc{eq} \label{eq_keyeq2},
\end{align}
where $\deltavin$ has been negated. This changes the LUT of the bootstrap in \autoref{table:LUTmin}, but the general techniques developed are still valid. As the constant near the $\pmb{M_{i-1,j-1}}$ term is now 2 instead of 4, we have an equivalent noise of:
\begin{align*}
   \epsilon^2_{key} = N_H \cdot \epsilon^2_{PBS}  + 4 \cdot N_M \cdot \epsilon^2_{PBS} + 9 \cdot N_L \cdot \epsilon^2_{PBS} + \epsilon^2_{PBS}.
\end{align*}

\begin{figure}[t!]
    \centering
    \includegraphics[width=0.4\textwidth]{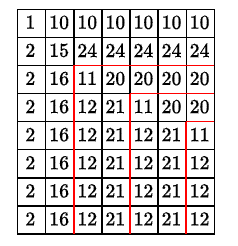}
    \caption{The relative value of the noise in each cell of the calculations. The red line indicates a refresh of the noise is needed using a bootstrap operation. This is for a parameter set that can handle up to 25 additions.}
    \label{fig:noise5}
\end{figure}

We now have improved noise equations. To make sure our noise does not surpass the threshold, we have to refresh the noise in the $\deltavin$ and $\deltavout$ terms just before the noise threshold is exceeded. \autoref{fig:noise5} depicts the noise in the ciphertexts, where the red line indicates where refreshing is needed. This example is for a parameter set that allows a maximum of 25 additions of ciphertexts before bootstrap is needed. 

In practice, the standard parameter sets used in TFHE-rs can typically handle more than $4\,000$ additions before a bootstrap is needed, according to our calculations. As such, only in edit distance calculations with very large words does one have to take these refreshes into account. We have experimentally verified this calculation by running the bootstrap on increasingly noisy ciphertexts and testing when an error occurs.

\subsection{Skipping irrelevant cells}
\label{sec:skippingcells}

When calculating the $d$-matrix (or equivalently, the horizontal and vertical differences $\deltah$ and $\deltav$), certain cells do not influence the final result and thus do not need to be computed.

This can be understood as follows:
\begin{itemize}
    \item \textbf{Maximum Levenshtein Distance}: The maximum possible value of the Levenshtein distance is bounded. In the worst case, it is $\max(m, n)$, meaning that every character in one string needs to be substituted to match the other string.
    \item \textbf{Shortest Path Analogy}: Computing the Levenshtein distance is analogous to finding the shortest path through the $d$-matrix, where the cost of each move is defined by \autoref{eq_wf}. Horizontal and vertical steps always have a cost of 1, while diagonal steps depend on the value of $\textsc{eq}$.
    \item \textbf{Path Cost Bounds}: The minimal cost of traversing from the top-left to the bottom-right corner of the $d$-matrix is $m + n$ (i.e., $m$ vertical steps and $n$ horizontal steps). This value exceeds the maximum possible Levenshtein distance, which implies that not all cells are relevant to the computation.
\end{itemize}

These observations reveal that many cells in the $d$-matrix are unnecessary for determining the final distance. Without loss of generality, consider the case where $m = n$. According to Ukkonen~\cite{Ukkonen85a}, if a cell that is $k$ steps away from the main diagonal is on the shortest path, the path has a minimum path cost of $2k$ (consisting of at least $k$ horizontal and $k$ vertical steps). 
Thus, cells located more than $\lfloor k/2 \rfloor$ steps from the diagonal can be excluded from computation without affecting the accuracy of the result.

In a more extreme situation, one can compute approximate Levenshtein distances, where the result is accurate up to a Levenshtein distance of $\ell$, but for larger Levenshtein distances, the output might be wrong (i.e., the output might be larger than expected). In this case, one only has to compute cells that are within $\lceil \ell /2 \rceil$ of the diagonal. This is useful when you want to determine if two strings are approximately equal. 
Concretely, when $m=n$, we do not need to calculate all the $m^2$ cells, but we can reduce this to
\begin{equation*}
\label{eq_skip}
   m+2\cdot \sum^\ell_{i=1}(m-i) = m \cdot (2 \ell + 1) - \ell^2 - \ell .
\end{equation*}

Furthermore, this modification precludes the calculation of the edit distance value from the final \deltah row. Consequently, the final score is determined by summing all \deltav and \deltah elements along the diagonal.

\subsection{The resulting algorithm} % And conclusion

Combining all of the above approaches leads to Algorithm~\vref{alg_levenshtein}. This algorithm has as inputs a matrix $\mathbf{\textsc{\textbf{eq}}_9}$ and parameter $\ell$. 
Matrix $\mathbf{\textsc{\textbf{eq}}_9}$ will contain the equality information for each character pair. That is, element $[i,j]$ will contain a 9 if $x_i = y_j$ and otherwise will contain 0. Parameter $\ell$ will denote the approximation level. By assigning $\ell=\lceil \max(m,n)/2 \rceil$, the exact edit distance will be calculated.

\begin{algorithm}[!ht]
  \caption{\LVS}
  \label{alg_levenshtein}
  \begin{algorithmic}[1]
    % \Require{Any of the edit operations has a unit-cost}
    
    \Statex
    \Function{Edit Distance}{$\mathbf{\textsc{\textbf{eq}}_9},\ell$}
        
        \Statex \(\triangleright\) Setup
        
        \Let{$h$}{OneMatrix$[0..m,0..n]$}
        \Let{$v$}{OneMatrix$[0..m,0..n]$}
        \Let{LUT$_{\text{min}}$[key]}{\autoref{table:LUTmin} }

        \item[]
        \Statex \(\triangleright\) Main Algorithm

        \For{$j \gets 1 \textrm{ to } n$}
            \For{$i \gets 1 \textrm{ to } m$}
                            \If{$|i-j| \leq \ell$}

                \Let{$key$}{$(1 - v[i,j-1]) + 3 \cdot (1 + h[i-1,j]) + \mathbf{\textsc{\textbf{eq}}_9}[i,j]$}
                \Let{$min$}{PBS($key$, LUT$_{\text{min}}$)}
                
                \Let{$v[i,j]$}{$min-h[i-1,j]$}
                \Let{$h[i,j]$}{$min-v[i,j-1]$}

                \EndIf
            \EndFor
        \EndFor
        \State \Return{$\Sigma_{i=1}^{m+1} h[i, i] + \Sigma_{i=0}^{n} v[i+1, i]$}
    \EndFunction
  \end{algorithmic}
\end{algorithm}

\begin{table}[]
\centering
\caption{Overview of the PBS load of the different algorithms for a single cell, using ASCII encoding. The exact number of programmable bootstrap needed for the WF and Myers algorithm is estimated here, as it depends on the exact scenario. More information can be found in \autoref{sec:results}}
\label{tab_cell_pbs}
\begin{tabular}{@{}lcccc@{}}
\toprule
Algorithm \; &  \; Levenshtein  \; &  \; Improvement factor \;  \\ \midrule
WF      & 94      & $1\times$                          \\
Myers         & 16    &  $5.87\times$   \\                     
Ours                                     & 1        & $94\times$                                              \\ \bottomrule
\end{tabular}
\end{table}

To make a fair comparison, we ran our WF and Myers implementations and counted the number of PBS performed. The WF and Myers algorithm would need on average, approximately 94 PBS and 16 PBS to calculate a cell, when using ASCII encoding. While the size of the other algorithms depends on the character encoding and length of the input strings, our algorithm is independent of these. The cost for our algorithm will remain constant, requiring only one PBS.

\section{Equality checking}
\label{sec:equality}

In the previous section, we described an algorithm that only uses 1 PBS per cell to calculate the Levenshtein distance. For each cell, one also has to calculate the equality between the corresponding characters of the input strings. This typically costs more than 1 PBS per cell, and thus the equality calculation is the most expensive operation of the full Levenshtein algorithm in our case. In this section, we improve the equality calculation in two ways: we introduce a technique that allows doubling the number of plaintext bits in one PBS equality operation, and then we propose a new technique to more efficiently look at larger symbols, notably 7-bit ASCII symbols. 

\subsection{Doubling the equality PBS size}
\label{sec:comparison}

The standard approach to equality checking involves dividing the binary representation of the input letters into chunks of 2-bit and then pairwise comparing the corresponding bits. This method is also implemented in the software library TFHE-rs~\cite{TFHE-rs}. For instance, when comparing a 2-bit $x$ with a 2-bit $y$, one computes the key $x + 4 \cdot y$ and then performs a PBS that maps $x=y$ to 1 and all other values to 0. Using the standard parameter size (2-bit plaintext, 2-bit carry), this method can only handle inputs of at most 2 bits. 

We present a new method that can handle 4-bit symbols. To compare two (4-bit) chunks $x$ and $y$, we subtract both values, resulting in a variable with a value between $-15$ and $15$, and a value of 0 if and only if both chunks are the same. As before, this results in more values than the normal 16-value PBS lookup. By choosing the following lookup table for the PBS:

\begin{equation}
    \text{LUT}_{\textsc{eq}} = 
    \begin{cases}
        1 & \text{if } (x-y) = 0 \\
        0 & \text{else},
    \end{cases}
    \label{eq_eqlut}
\end{equation}

we can have a 31-value lookup due to the negacyclic property of the lookup and the abundance of 0 values. 
Note that in our Levenshtein calculation, we sometimes want to calculate $\textsc{eq}_9$, for which we adapt the LUT to:

\begin{equation}
    \text{LUT}_{\textsc{eq}_9} = 
    \begin{cases}
        9 & \text{if } (x-y) = 0 \\
        0 & \text{else}.
    \end{cases}
    \label{eq_eqlut9}
\end{equation}

Thus, our new method of equality checking can handle symbols of double size. For typical large integers where the integer is divided into ciphertexts that each contains 2-bit chunks, two 2-bit equality checks can be combined into one by calculating $x = 4 \cdot x^{(2)} + x^{(1)}$ (and similarly for $y$), thus halving the PBS cost. This also means that an equality check for larger inputs can be done at half of the PBS cost using our method.

\subsection{Equality check for large-sized symbols}

Larger characters are typically divided into smaller symbols of size $t$, usually 2 to 4 bits. A sub-equality operation is performed for each pair of chunks, after which the results of these sub-equality operations are combined to produce the final equality result.

The TFHE-rs library provides a two-step algorithm for calculating equality. First, corresponding chunks are compared using the method $x + (y \ll 2)$, as described earlier. This will produce a sub-equality that will compare 2-bit or character data. The ciphertext will contain a one if the two parts are equal, a zero in the other case.

In the second step, the sub-equality results are summed together, in a triangular way, to find the overall equality. All of the sub-equalities are divided into groups of maximally $t-1$ elements. All sub-equalities in a group are summed together and a PBS is applied to the sum to check if the sum reaches its maximum possible value, i.e. $t-1$. The results are now again grouped into maximally $t-1$ elements, summed together and used in a PBS. 
This process is repeated until a single ciphertext is obtained, representing the result of the equality check. If all of the sub-equalities of step one consist of a one, the final equality result will also depict a one. Algorithm~\ref{alg_large_cmp} outlines this approach for characters of at most 30-bit.

For example, when comparing two ASCII characters (7-bit), each character is encoded into four ciphertexts. After computing the sub-equalities, the four outcomes are summed, and a PBS is used to verify if the total sum equals 4. 

This method can be further improved using a hybrid approach of our custom equality subcheck and the TFHE-rs combination phase.

In this hybrid approach, characters are encoded into consecutive 4-bit chunks. In the first part, subcomponents are computed using our subtraction method. In the second part, the TFHE-rs aggregation step is used to combine the subcomponents efficiently. This reduces the cost of calculation by roughly half.

\begin{algorithm}[!htb]
  \begin{algorithmic}[1]
  \Require{Two encoded characters $x$ and $y$}
  \Require{$x, y$ are split into $n$ encrypted symbols, $n<16$, each symbol has $t=2^p$ size plaintext}
  \Statex \(\triangleright\) Equal Part
    \For{$i \gets 0 \textrm{ to } n$}
        \Let{$z_0$}{$x^{(i)} + 4\cdot y^{(i)}$} 
        \Let{$\textsc{eq}^{(i)}$}{PBS($z_0;$ LUT$_\textsc{eq}$)}
    \EndFor
    \item[]
    \Statex \(\triangleright\) Merging Part

    \Let{Acc}{0}
    \For{$i \gets 0 \textrm{ to } n$}
        \Let{Acc}{Acc + $\textsc{eq}^{(i)}$} 
    \EndFor
    \Let{Acc}{PBS(Acc, LUT$_{\text{max}}$)} 
    \State \Return{Acc}
  \end{algorithmic}
  \caption{TFHE-rs compare method for characters with maximum size of 30 bit.}
  \label{alg_large_cmp}
\end{algorithm}

\subsection{Equality check for medium-sized symbols (e.g., ASCII)}

While this combination method of TFHE-rs is efficient for large input symbols, we propose a better method for medium-sized inputs. Specifically, our method outperforms state-of-the-art for 5- to 16-bit inputs, specifically for ASCII inputs. For this explanation, we will assume 7-bit input characters and a 4-bit combined message and carry space.

The first step in our method is to decide on the representation of our characters. We will encode our characters using a 4-bit and a 3-bit symbol, each encrypted in one ciphertext. We will then perform the equality check between the 4-bit symbols as explained in the previous section. The result is a single bit $\textsc{eq}$ denoting equality, which is combined with the 3-bit symbol in the following way:

\begin{align*}
  2 \cdot (x^{(2)}_i - y^{(2)}_i) + (1-\textsc{eq}^{(1)}).
\end{align*}

The result of this linear computation is 0 if and only if both the 4-bit symbols and the 3-bit symbols are the same. Thus, we can perform the same PBS lookup as before on this result to calculate the equality of the characters. The formal representation is given in Algorithm~\ref{alg_ascii}.

\begin{algorithm}[!htb]
  \begin{algorithmic}[1]
  \Require{Two ASCII encoded characters $x$ and $y$}
  \Require{$x, y$ are split into 4-bit symbols $(x^{(1)}, y^{(1)})$ and 3-bit symbols $(x^{(2)}, y^{(2)})$ }
   
        \Let{$z_0$}{$x^{(1)} - y^{(1)}$} \Comment{first 4 bits of the character}
        \Let{$\textsc{eq}^{(1)}$}{PBS($z_0;$ LUT$_\textsc{eq}$)}
        \Let{$z_1$}{$2 \cdot (x^{(2)} - y^{(2)}) + (1 - \textsc{eq}^{(1)})$} \Comment{last 3 bits of the character}
        \Let{$\textsc{eq}$}{PBS($z_1;$ LUT$_\textsc{eq})$}

    \State \Return{$\textsc{eq}$}
  \end{algorithmic}
  \caption{Our equality check for 7-bit characters, split into 4 and 3-bit}
  \label{alg_ascii}
\end{algorithm}

The result is that we can perform an ASCII equality check with two PBS, while other methods would need five lookups when using state-of-the-art 2-bit equality techniques or three lookups when using our 4-bit equality technique. 

\subsection{Conclusion}

\begin{figure}[!ht]
    \centering
    \includegraphics[width=\textwidth]{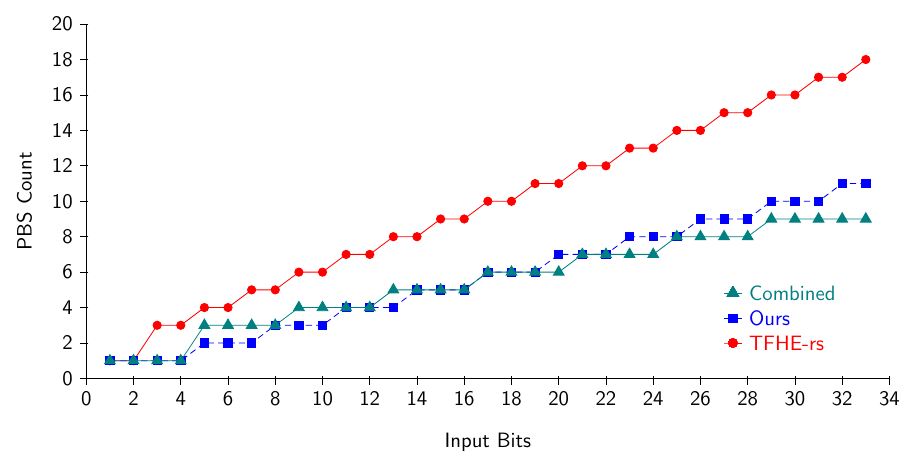}
    \caption{Comparison of the PBS count for the three equality calculation techniques. }
    \label{fig_eq}
\end{figure}

To calculate the equality between two characters, there are three options: TFHE-rs, our own, or the combined approach. All three approaches demonstrate different PBS behaviour. Figure~\ref{fig_eq} denotes the PBS loads as a function of the input size. From this analysis, we can see that our method is the best for characters of up to 16-bit, which is ideal for an ASCII usecase. For larger characters, the combined method performs better.
Furthermore, our method reduces the memory footprint of the encrypted character by up to a factor of two.

\section{Preprocessing}
\label{sec:preprocessing}

Even with reduced cost for the equality check as discussed in the previous section, the equality checking is typically still more expensive than the main algorithm. More specifically the main algorithm uses 1 PBS per cell, while the equality calculation costs $C$ PBS calculations per cell with $C$ a constant depending on the input characters. In this section we will show that in the specific case that one of the strings is unencrypted, we can perform a preprocessing to speed up the equality calculations, making their total cost linear in the input size (i.e. $C \cdot |S| \cdot m$, for $m$ the size of the encrypted input). 

The idea of this preprocessing is to precompute the output of the equality for each possible character of the alphabet for the encrypted input (which has cost $|\Sigma|\cdot m$), similar to a technique proposed by Myers~\cite{Myers99} to allow efficient bitsliced implementations. During the equality check one then can perform a simple lookup using the unencrypted input character to find the result of the equality, which does not require any PBS.

More specifically, during the preprocessing phase, we check the equality of each character in the encrypted string with every possible character in the alphabet. The results are then stored in an encrypted table and when the edit distance is calculated, we can use the unencrypted data to obtain the encrypted equality information. 

For instance, if the word 'abbey' is the encrypted string and only encodes lowercase letters, we would store $5\times 26$ elements, as shown in Table~\ref{tab_pre}. Note that it is also possible to store $\textsc{eq}_9 = \text{enc(9)}$ and can therefore be used directly in the key calculation.

\begin{table}[!ht]
\centering
\caption{\label{tab_pre} Preprocessed storage of the word `abbey' using only lowercase letters.}
\begin{tabular}{@{}cccccc@{}}
\toprule
character & 1                                  & 2                                  & 3                                  & 4                                  & 5                                  \\ \midrule
a         & \textbf{enc(1)} & enc(0)                             & enc(0)                             & enc(0)                             & enc(0)                             \\
b         & enc(0)                             &  \textbf{enc(1)} & \textbf{enc(1)} & enc(0)                             & enc(0)                             \\
c         & enc(0)                             & enc(0)                             & enc(0)                             & enc(0)                             & enc(0)                             \\
d         & enc(0)                             & enc(0)                             & enc(0)                             & enc(0)                             & enc(0)                             \\
e         & enc(0)                             & enc(0)                             & enc(0)                             & \textbf{enc(1)} & enc(0)                             \\
...       & ...                                & ...                                & ...                                & ...                                & ...                                \\
y         & enc(0)                             & enc(0)                             & enc(0)                             & enc(0)                             & \textbf{enc(1)} \\
...       & ...                                & ...                                & ...                                & ...                                & ...                                \\ \bottomrule
\end{tabular}
\end{table}

The preprocessing step has a cost of $|\Sigma| \cdot m$, where $m$ is the length of the encrypted string and $|\Sigma|$ is the size of the character set. This is specifically useful in the case of matching DNA sequences, we would only need to store the 4 characters.
Moreover, if during encryption we know we will only deal with a specific subset of characters $S \subset \Sigma$, we can reduce the cost to $S \cdot m$ by constructing a table only for the characters in $S$. As an example, this could be the case with names where characters like Y and Q might not occur in the unencrypted query string.

A standard edit distance calculation requires $m \cdot n$ equality operations. Therefore, if $|\Sigma| < n$ (or $S < n$), this optimisation becomes advantageous. For example, with full ASCII, the method becomes more efficient for $n > 128$. For lowercase letters, it applies when $n > 26$, and for DNA sequences, it becomes beneficial when $n > 4$. 

This approach is particularly useful when matching a plaintext string against a database. In the case where the database is unencrypted and the query is encrypted, one needs to do the query only one time and the result can be used for multiple lookups. If the database is encrypted and the query is not encrypted, one can preprocessing each encrypted string in the database in an offline preparation phase. This means that  future plaintext equality checks can be done without the need for additional PBS. Once a new plaintext string needs to be matched, one only needs to perform a lookup in the table.

\section{Results}
\label{sec:results}

In this section, we will compare our \LVS{} algorithm to the state-of-the-art in FHE edit distance calculations. A challenge in analysing the efficiency of our improvements is that there is only one implementation of the Levenshtein distance available in TFHE, which is an implementation that is used as a demonstrator for the concrete compiler~\cite{Concrete}. To have more comparison points, we implemented standard versions of the Wagner-Fischer and Myers algorithms using the TFHE-rs library. Both algorithms rely on standard 2-bit message and 2-bit carry ciphertexts as fundamental building blocks. In our experiments, we used ASCII encoding to encrypt each character. The Wagner-Fischer and Myers algorithms are implemented using the state-of-the-art equality check techniques as available in the TFHE-rs library.

Our algorithm uses both our improved Levenshtein (algorithm~\ref{alg_levenshtein}) and our improved equality check (algorithm~\ref{alg_ascii}). A complete overview of the algorithm is given in Algorithm~\vref{algo_complete}.
We discern four versions of our algorithm in our experiments: 
\begin{itemize}
    \item \textbf{\LVS{} Exact}: Full calculation of all the cells in the edit distance algorithm.
    \item \textbf{\LVS{} Exact Skipping}: Calculation of all the cells that are relevant for the end result, but skipping irrelevant cells as discussed in \autoref{sec:skippingcells}.
    \item \textbf{\LVS{} Approx. $\pmb{\ell = n/4}$}: Calculation of the approximate edit distance that is accurate for distances lower than $m/4$ as discussed in \autoref{sec:skippingcells}. To simplify the discussion we assume $m=n$ for the approximate results.
    \item \textbf{\LVS{} Approx. $\pmb{\ell = 10}$}: Calculation of the approximate edit distance that is accurate for distances lower than $10$ as discussed in \autoref{sec:skippingcells}. 
\end{itemize}

\subsection{Counting the bootstraps}

In this section, we analyse the theoretical cost of the algorithms, focusing on the number of bootstraps required.
Due to the insertion of auxiliary PBS operations for message-carry maintenance in the TFHE-rs library, determining the exact number of PBS used in theory is not trivial. We experimentally determined the average number of PBS per cell in the WF and Myers algorithms by running the program and counting the PBS. For the \LVS{} algorithm, we derive a first-order estimate based on the bootstraps incurred by non-linear operations within each cell, excluding incidental bookkeeping bootstraps introduced for noise management.

\begin{table}[!ht]
\centering
\caption{Overview of the PBS load of the different algorithm to match an $n$-element with an $m$-element string. The preprocessing and Levenshtein column denote the operations per cell, the last column is a first-order approximation of the total cost over all cells.}
\label{tab_pbs}
\begin{tabular}{@{}lccr@{}}
\toprule
Algorithm & \multicolumn{1}{l}{Preprocessing} & \multicolumn{1}{l}{Levenshtein} & \multicolumn{1}{r}{Total}    \\ \midrule
WF        & 5                                 & 28                              & $33\cdot mn$                    \\
Myers     & 5                & 13                              & $18\cdot mn$ \\ \midrule
\LVS{} Exact      & 2                                 & 1                               & $3\cdot mn$                                \\ 
\LVS{} Exact Skipping  & 2                              & 1                               & $\approx 2.25 \cdot mn$                       \\ \hline
\LVS{} Approx. $\ell = m/4$ & 2                                 & 1                           & $ \approx 6 \cdot\ell\cdot  n$        \\
\LVS{} Approx. $\ell = 10$ & 2                                 & 1                           & $ \approx 60 \cdot n$      \\\bottomrule
\end{tabular}
\end{table}

From \autoref{tab_pbs} one can see that our algorithm significantly reduces the number of required bootstraps to compute a cell in the encrypted domain. Firstly, we need only $2.5\times$ fewer PBS for equality calculations. Secondly, we achieve a reduction of $94\times$ and $16\times$ less PBS compared to the Wagner-Fischer and Myers algorithms, respectively. Overall, for the full distance calculation, our best exact method demonstrates a $44\times$ reduction in the number of PBS required over the Wagner-Fischer algorithm and an $9.33\times$ improvement over the Myers algorithm.

\subsection{Implementation results}

We implemented the Wagner-Fischer, Myers, and our custom algorithm using Rust v1.80.0, TFHE-rs v0.7.2~\cite{TFHE-rs} and concrete compiler v2.8.1~\cite{Concrete} on an Ubuntu 22.04 system. All experiments were conducted on a dual AMD EPYC 9174F 16-Core Processor (for a total of 64 threads). The parameter set used for the Wagner-Fischer and Myers implementations are 2-bit message and 2-bit carry ciphertext which is the most standard choice., For the \LVS{} implementation, we opted not to use the carry space, instead employing 4-bit message 0-bit carry parameter set. This implementation choice makes the implementation easier, but has little effect on the performance. All timing results represent the full calculation of the edit distance, including both preprocessing and the main algorithm. Each experiment was conducted using the sequential implementation of the shortint API. We will discuss parallelisation options for these algorithms in next section.

\begin{table}[!ht]
\centering
\begin{minipage}{\textwidth}
 
\caption{Latency results in seconds of the calculation of the edit distance, using ASCII encoding, in the encrypted domain. }
\label{tab_lat}
 
\begin{tabular*}{\textwidth}{@{\extracolsep{\fill}}@{}llHrllHrllHr@{}}
\toprule
  & \multicolumn{3}{c}{$m=8$}     &  & \multicolumn{3}{c}{$m=100$}     &  & \multicolumn{3}{c}{$m=256$}      \\ \cmidrule(lr){2-4} \cmidrule(lr){6-8} \cmidrule(lr){10-12} 
    & \multicolumn{1}{c}{Seq.} & & & & \multicolumn{1}{c}{Seq.} & & & & \multicolumn{1}{c}{Seq.} &  \\ \midrule
\cite{CheonKL15}\footnote{Using the DGHV scheme, an 80-bit security level and other hardware.} & 27.54         &              &  &                &               &  &               & &   & &             \\
\cite{Concrete}                                                                                  & 241.10        &              & $1\times$ &  & 12h 36m               &              & $1\times$ &  & 6d 21h\footnote{Extrapolated based on 24300 cell calculations.} &                  & $1\times$            \\ % 45361.6569 and 220722.6776 (for n=24300) -> 595279.068
WF                                                                                             & 77.59         & 32.28         & $3.1\times$ & & 3h 24m      & 1h 29m    & $3.7\times$ & & 22h 19m      & 9h 8m & $7\times$   \\
Myers                                                                                          & 17.81         & 12.00         & $14\times$ & & 38m 12s         & 15m 28s       & $20\times$& & 4h 7m         & 1h 41m    & $40\times$   \\ \midrule
\LVSShort Exact                                                                                         & 2.83          &             & $85\times$& & 7m 19s          &              & $103\times$ && 48m 23s          &             & $205\times$   \\
\LVSShort Exact Skip                                                                                    & \textbf{2.28} &              & $106\times$ & & \textbf{5m 35s} &              & $135\times$ & & \textbf{35m 41s} &              & $278\times$   \\ \midrule \midrule
\LVSShort Apx. {\footnotesize$\ell=\frac{m}{4}$}                                                                                  & 1.50          &             & $161\times$ & & 3m 19s          &              & $228\times$ & & 20m 52s          &            & $475\times$   \\
\LVSShort Apx. {\footnotesize$\ell=10$} \;                                                                                  &           &              & & & 1m 26s          &              & $527\times$ && 3m 50s          &              & $2588\times$   \\ \bottomrule

\end{tabular*}
 
\end{minipage}
\end{table}

The results in this table demonstrate that our \LVS{} algorithm significantly outperforms state-of-the-art TFHE algorithms. Specifically, we achieve up to $278\times$ speedup over the best available algorithm, and $39\times$ speedup over our Wagner-Fisher implementation. 

While we have included the numbers for edit distance for second generation FHE, these results are run with different hardware and lower security settings, and are therefore not suitable for a fair comparison. 

Additionally, skipping irrelevant cells as discussed in \autoref{sec:skippingcells} is an efficient way to optimise the algorithm by $25\%$ for exact calculations, and limiting the accuracy for large edit distances can give another significant efficiency improvement. These results clearly show that our algorithm consistently outperforms existing approaches across all scenarios.

\subsubsection{Parallelism: } Previous results were achieved using the sequential TFHE-rs parameter set. Greater efficiency can be attained by developing parallel implementations. While a comprehensive investigation into parallelisation is left for future work, we will provide some preliminary insights and suggestions here. 

In general one can think of three levels at which to parallelise:
\begin{itemize}
    \item \textit{Parallelisation over operations}: Some operations require multiple PBS operations that can be performed in parallel. This is the case for for example additions or a min function over larger integers (as used in Wagner-Fisher) or Boolean operations on multiple inputs at the same time (as used in Myers). TFHE-rs has a parallel implementation available for these operations. This strategy can not be used for our method, as we only have 1 PBS per cell of the alogrithm. 
    \item \textit{Parallelisation over cells}: The edit distance algorithm allows for parallelisation across multiple cells that are independent of one another. Specifically, cells equidistant from the first cell can be computed simultaneously. This approach enables efficient parallel computation for the majority of the cells, with only the first and last cells experiencing limited parallelisation. For large string sizes, this strategy achieves near-complete parallelisation, significantly enhancing execution efficiency. The implementation of this parallelisation method is deferred to future work.
    \item \textit{Batch inputs matching}: In case of multiple edit distance calculations that need to be performed at the same time, one can easily parallelise the calculation by performing all lookups in parallel. This is especially relevant to parallellise the workload for database lookup scenarios. Table~\ref{table_results_parallel} shows a comparison of the algorithms under parallel computations.
\end{itemize}

\begin{table}[!ht]
\centering
\caption{Latency in seconds of parallelised implementation using 64 threads each calculating one Levenshtein distance. The first results of WF and Myers in the table only use the parallel TFHE-rs API, the following results use both the parallel API and batched inputs. Relative speedups compare with sequential results from \autoref{tab_lat}.}
\label{table_results_parallel}
 
\begin{tabular*}{\textwidth}{@{\extracolsep{\fill}}lllllll}
\toprule
             & \multicolumn{2}{c}{$m=8$} & \multicolumn{2}{c}{$m=100$} & \multicolumn{2}{c}{$m=256$} \\ \cmidrule(lr){2-3} \cmidrule(lr){4-5} \cmidrule(lr){6-7} 
             & Lat.   &         & Lat.    &         & Lat.   &         \\ \midrule
WF (TFHE-rs parallel ops.)          & 32.28       &  $2.4\times$       & 1h 29m     &   $2.4\times$   & 9h 8m    &   $2.4\times$      \\ 
Myers (TFHE-rs parallel ops.)       & 12.00       & $1.5\times$   & 15m 28s     & $2.5\times$   & 1h 41m    & $2.4\times$   \\ \midrule
WF (batch inputs)          & 2.60       &  $29.8\times$       & 6m 30s     &  $31.4\times$       & 42m 52s    &    $31.2\times$     \\ 
Myers (batch inputs)       & 0.91       & $19.5\times$   & 2m 45s     & $13.9\times$   & 18m 30s    & $13.4\times$   \\ \midrule
\LVSShort Exact        & 0.15       & $18.9\times$     & 14.7       & $29.9\times$   & 1m 35       & $30.5\times$   \\
\LVSShort Exact Skip   & 0.14       & $16.3\times$   & 11.3       & $29.6\times$   & 1m 12       & $29.7\times$   \\
\LVSShort Exact $\ell = \frac{m}{4}$ & 0.11       & $13.6\times$   & 6.88       & $28.9\times$   & 43.7       & $28.6\times$   \\
\LVSShort Exact $\ell =10$ & -          & -        & 3.39       & $25.4\times$    & 10.4       & $22.1\times$    \\ \bottomrule
\end{tabular*}
\label{table:results_parallel}
\end{table}

Execution using the TFHE-rs parallel API shows only a modest impact on execution time (approximately $2.4\times$), highlighting the need to explore alternative sources of parallelism. In contrast, our batched inputs significantly accelerate computation, achieving near-optimal speedup of $32\times$, which aligns with full core utilisation of our CPU. This demonstrates that batching is an effective strategy for maximising parallelism and fully utilising server resources. However, for single lookups, alternative parallelisation approaches, such as parallelisation over cells, will need to be explored. 

In certain cases, such as the Myers approach, the optimal $32\times$ speed-up is not fully realised. This is due to memory management factors, including cache saturation and reliance on RAM, which could be addressed with careful memory handling.

\subsubsection{Preprocessing: }

In \autoref{sec:preprocessing} we discussed preprocessing when one of the strings is not encrypted. Table~\ref{tab_unenc} compares execution of the algorithm with and without preprocessing, based on the scenarios outlined above. All of the calculations are done using our \LVS{} algorithm.

\begin{table}[!ht]
\centering
\caption{Latency results for the case of one unencrypted string, both the latency of the building up the preprocessing table and main algorithm are given. All results in seconds. The relative speedup is given for the main calculation.}
\label{tab_unenc}
\begin{tabular*}{\textwidth}{@{\extracolsep{\fill}}llllllllll@{}}
\toprule
              & \multicolumn{3}{c}{$n=8$}                           & \multicolumn{3}{c}{$n=100$}                         & \multicolumn{3}{c}{$n=256$}                        \\ \cmidrule(lr){2-4} \cmidrule(lr){5-7} \cmidrule(l){8-10} 
              & \multicolumn{1}{c}{pre} & \multicolumn{1}{c}{main} &  & \multicolumn{1}{c}{pre} & \multicolumn{1}{c}{main} &  & \multicolumn{1}{c}{pre} & \multicolumn{1}{c}{main} \\ \midrule
\LVSShort Exact - No Prep. & \multicolumn{1}{c}{-} & 2.83 & $1\times$ & \multicolumn{1}{c}{-} & 439 & $1\times$ & \multicolumn{1}{c}{-} & 2903 & $1\times$ \\
\LVSShort Exact & 29.6 & 0.93 & $3\times$ & 369 & 146 & $3\times$ & 942 & 950 & $3\times$ \\ \midrule
\LVSShort Exact Skip  - No Prep.& \multicolumn{1}{c}{-} & 2.28 & $1\times$ & \multicolumn{1}{c}{-} & 335 & $1\times$ & \multicolumn{1}{c}{-} & 2141 & $1\times$ \\
\LVSShort Exact Skip & 29.5 & 0.75 & $3\times$ & 367 & 110 & $3\times$ & 946 & 717 & $3\times$ \\ \midrule
\LVSShort Exact $\ell = \frac{m}{4}$  - No Prep. & \multicolumn{1}{c}{-} & 1.5 & $1\times$ & \multicolumn{1}{c}{-} & 199 & $1\times$ & \multicolumn{1}{c}{-} & 1252 & $1\times$ \\
\LVSShort Exact $\ell = \frac{m}{4}$ & 29.6 & 0.49 & $3\times$ & 368 & 65 & $3\times$ & 946 & 420 & $3\times$ \\ \midrule
\LVSShort Exact $\ell = 10$  - No Prep. & \multicolumn{1}{c}{-} &  &  & \multicolumn{1}{c}{-} & 86 & $1\times$ & \multicolumn{1}{c}{-} & 230 & $1\times$ \\
\LVSShort Exact $\ell =10$ &  &  &  & 370 & 29 & $3\times$ & 941 & 77 & $3\times$ \\
\bottomrule
\end{tabular*}
\end{table}
From this table we can see that preprocessing reduces the computation with a factor $3\times$, as can be expected from the fact that character equality costs 2 PBS per cell while the Levenshtein calculation costs 1 PBS per cell. Note that the preprocessing numbers can be improved more if some ASCII characters are not used. 

In general, one can discern three scenarios where preprocessing is useful:

\begin{itemize}
    \item \textit{Single Lookup}: In this scenario, a speedup is achievable when the alphabet size is smaller than the string length (i.e., $|\Sigma| < n$), as discussed in \autoref{sec:preprocessing}. From the table, one can observe this effect for $n=256$, where the total cost of preprocessing plus the main algorithm is lower than the cost without preprocessing. 
    \item \textit{Encrypted query against a large unencrypted database}: In this scenario, preprocessing needs to be performed only once. After that, only the main algorithm is executed. For large datasets, this approach clearly demonstrates a speedup approaching a factor of $3\times$. 
    \item \textit{Unencrypted query against an encrypted database:} Here, the database can undergo a one-time preprocessing step, either in plaintext or in the encrypted domain. Following this, only the main computation cost is incurred, resulting in a similar speedup of $3\times$. 
\end{itemize}

\subsection{Comparison with MPC implementations}

In Multi-Party Computation (MPC) settings, a trade-off exists between computation time and bandwidth. 
Both FHE and MPC enable privacy-preserving computations, but directly comparing their specific security guarantees is challenging due to their differing threat models and trust assumptions. MPC categorises adversarial models into two main types: passive MPC, also called the \emph{honest-but-curious} model, which operates under the assumption that compromised parties will adhere to the protocol while trying to infer confidential information. On the other hand, active MPC is designed to withstand adversaries who may deviate from the protocol in arbitrary ways to compromise correctness or obtain sensitive data.
\\ \\
Vanegas et al.~\cite{VanegasCA23} implemented edit distance calculation in an MPC setting, optimising their implementation for DNA strings (i.e., 2-bit characters). They considered an passive and active security model with LAN network delays for both Garbled Circuit (GC) and secret-sharing schemes operating over a $\mathbb{Z}_{2^k}$ domain.

\begin{table}[!ht]
\centering
\caption{Evaluation of the edit distance calculations using LAN setting MPC and FHE solutions for matching a 210-length DNA string. All tested on an \texttt{c6a.4xlarge} AWS EC2 instance.}
\label{tab_mpc}
 
\begin{tabular}{@{}cclrr@{}}
\toprule
Situation & Security & Algorithm & Lat. [s] & Data sent [MB] \\ \midrule
 &  & Yao's GC & 2.7 & 345.8 \\
 & \multirow{-2}{*}{Passive} & Semi$2^k$ & 103.0 & 113.1 \\ \cmidrule(l){2-5} 
 & & BMR-MASCOT & 9, 034.0 & 2.09 x 106 \\
\multirow{-4}{*}{\begin{tabular}[c]{@{}l@{}}MPC\\ \cite{VanegasCA23}\end{tabular}} & \multirow{-2}{*}{Active} & SPD$\mathbb{Z}_{2^k}$ & 368.5 & 14,893.8 \\ \midrule
FHE & - & \LVSShort Exact Skip & 1820.8 & 1.38 \\ \bottomrule

\end{tabular}%
 
\end{table}

For a fair comparison, we evaluated our algorithm using DNA strings of 210 characters in length on the same AWS EC2 instance of type \texttt{c6a.4xlarge}\footnote{\href{https://aws.amazon.com/ec2/instance-types/c6a/}{https://aws.amazon.com/ec2/instance-types/c6a/}}. Specifically, only one PBS was utilised for the equality calculation, leaving the remainder of the algorithm unchanged. In addition, our solution easily generalises to arbitrary character sets and is not restricted to domain-specific inputs.

As shown in Table~\ref{tab_mpc}, compared to the passive MPC case, our solution is $674\times$ slower but requires $251\times$ less data to be sent.
Compared to the active case, our solution is only $4.94\times$ slower but requires $10\;792\times$ less data.

In other experiments, Vanegas et al. \cite{VanegasCA23} calculate the edit distance for DNA strings of length 1020. All implementations considered (plaintext, FHE, and MPC) exhibit quadratic cost growth as the input string lengths increase. This trend is evident in our results for string lengths between 100 and 256 and is expected to hold for even larger input sizes. Consequently, the cost scaling between FHE and MPC remains independent of input length, meaning that the ratio of their computational costs remains approximately constant, regardless of how long the input strings are.
\\ \\
It is also worth noting that the approximate algorithm demonstrates significantly better performance due to its linear complexity. However, since this linear behaviour is consistent across all implementations (plaintext, FHE, and MPC), it does not affect the relative cost comparison between them.

\section{Conclusion}

This paper introduces a novel method for efficiently computing the edit distance on encrypted data within the TFHE framework. Our first contribution demonstrates how to streamline edit distance calculations by employing a compact ternary representation, reusing programmable bootstrapping (PBS) results, and computing a three-input minimum function in a single lookup. The resulting \LVS{} algorithm achieves a $94\times$ reduction in the number of PBS operations compared to traditional approaches.

Our second contribution enhances equality checks, particularly for medium-sized inputs. For ASCII encoding, we reduced the lookup cost from 5 PBS operations in the state-of-the-art to just 2 PBS operations.

Finally, we introduced a preprocessing stage that precalculates equality checks when one of the inputs is unencrypted, enabling an additional speedup. For ASCII inputs, this approach achieves up to a $3\times$ improvement.

Our implementation results demonstrate that the \LVS{} algorithm delivers speedups of up to $278\times$ over the best available implementation, underscoring its efficiency. These optimisations significantly advance the practicality of encrypted edit distance computations, reducing computational overhead and enhancing scalability for real-world applications.

\section*{Acknowledgements}
This work was supported in part by the Horizon 2020 ERC Advanced Grant (101020005 Belfort) and the CyberSecurity Research Flanders with reference number VOEWICS02. Wouter Legiest is funded by FWO (Research Foundation – Flanders) as Strategic Basic (SB) PhD fellow (project number 1S57125N).
\begin{figure}[h]
\centering
\includegraphics[width=0.6\columnwidth]{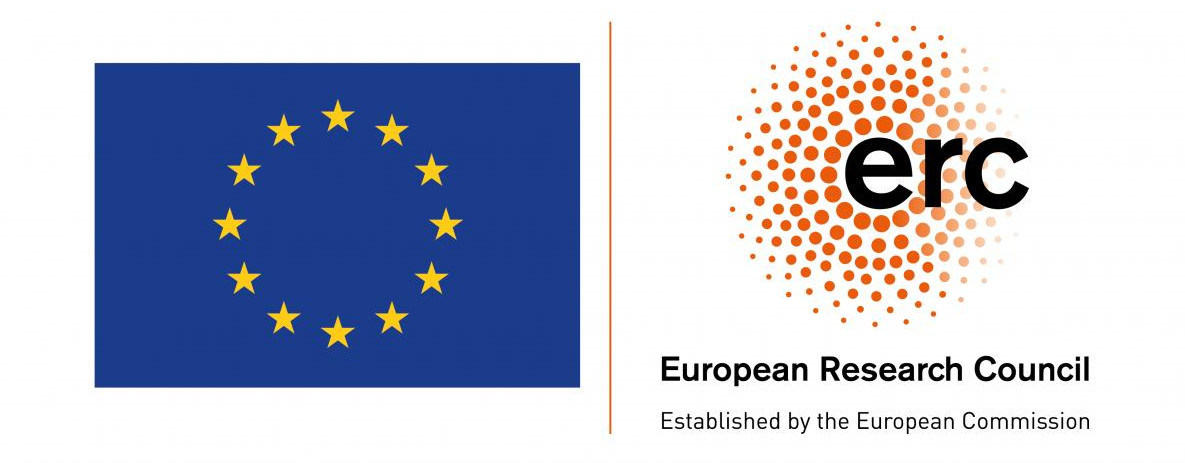}
\label{erc-eu-logos}
\end{figure}

\newpage

\bibliographystyle{alpha}
\bibliography{bib/abbrev0,bib/crypto,bib/non_crypto}

\iffalse
\fi

\section{Our complete algorithm}

\begin{algorithm}[H]
  \caption{Complete ASCII-based Encrypted Levenshtein}
  \label{algo_complete}
  \begin{algorithmic}[1]
    
    \Statex
    \Function{Edit Distance}{$x_{1..m}, y_{1..n}, \ell$}
        
        \Statex \(\triangleright\) Setup
        
        \Let{$h$}{OneMatrix$[0..m,0..n]$}
        \Let{$v$}{OneMatrix$[0..m,0..n]$}
        \Let{LUT-$\textsc{eq}_9$}{$[9,0,0,\dots]$}
        \Let{LUT-$\textsc{eq}$}{$[1,0,0,\dots]$}
        \Let{LUT-min[key]}{\autoref{table:LUTmin} }

        \For{$i \gets 1 \textrm{ to } m$}
            \Let{$v[i,0]$}{$1$}
        \EndFor
        \For{$j \gets 1 \textrm{ to } n$}
            \Let{$h[0,j]$}{$1$}
        \EndFor

        \item[]
        \Statex \(\triangleright\) Main Algorithm

        \For{$j \gets 1 \textrm{ to } n$}
            \For{$i \gets 1 \textrm{ to } m$}
                \If{$|i-j| < \ell$}
                \Let{$z_1$}{$x_{i}^{(1)}-y_{j}^{(1)}$}

                \Let{$\textsc{eq}_1$}{PBS($z_1;$ LUT-$\textsc{eq}$)}
                \Let{$\textsc{eq}_1$}{$1-\textsc{eq}_1$}
                
                \Let{$z_2$}{$x_{i}^{(2)}-y_{j}^{(2)}$}
                \Let{$z_2$}{$2\cdot z_2+\textsc{eq}_1$}
                \Let{$\textsc{eq}_9$}{PBS($z_2;$ LUT-$\textsc{eq}_9$)}

                \Let{$H_{in}$}{$h[i-1,j] + 1$}
                \Let{$V_{in}$}{$v[i,j-1] + 1$}

                \Let{$key$}{$(1 - v[i,j-1]) + 3 \cdot (1 + h[i-1,j]) + \textsc{eq}_9[i,j]$}
                \Let{$min$}{PBS($key$, LUT-min)}
                
                \Let{$v[i,j]$}{$min-h[i-1,j]$}
                \Let{$h[i,j]$}{$min-v[i,j-1]$}
                
            \EndIf
            \EndFor
        \EndFor
        \State \Return{$m+\Sigma_{i=0}^{n} h[m, i]$} 
    \EndFunction
  \end{algorithmic}
\end{algorithm}

\end{document}